\documentclass[preprint]{aastex}
\slugcomment{2020.6}

\shorttitle{Comments on the radiation-quiet anti-glitch} \shortauthors{H.Tong}

\begin{document}

\title{On the radiation-loud and radiation-quiet anti-glitches in magnetar 1E 2259+586}

\author{H. Tong\altaffilmark{1}}

\altaffiltext{1}{School of Physics and Electronic Engineering, Guangzhou University, Guangzhou 510006, China;\\ htong\_2005@163.com}

\begin{abstract}
The radiation-loud and radiation-quiet anti-glitches in magnetar 1E 2259+586 are discussed together. It is shown that the 2019 radiation-quiet anti-glitch in 1E 2259+586 is not constraining, because its amplitude is 2.5 smaller than the previous radiation-loud anti-glitch. We note that magnetars with high X-ray luminosity and low rotational energy loss rate are more likely to show anti-glitches. Anti-glitchers and radio-emitters of magnetars occupy opposite corner on the luminosity vs. rotational energy loss rate diagram. By introducing the X-ray luminosity parameter and the activity parameter, we try to characterize this trend quantitatively. A general picture of magnetar anti-glitch is given.
\end{abstract}


\section{Introduction}

Glitches are interesting timing anomalies of pulsars (Lyne \& Grham-Smith 2012). Almost all the glitches of normal pulsars are radiation-quiet, i.e. they are not accompanied by large amplitude increase of radiations in X-ray or other wavelength. However, this is not the case of magnetars. Some of the glitches in magnetars are accompanied by radiative events, while others are not (Dib \& Kaspi 2014). One the other hand, all radiative events seem to be accompanied by some kind of timing events (glitch, anti-glitch etc, Dib \& Kaspi 2014). The two magnetar-like activities in PSR J1846-0258 and PSR J1119-6127 (Gavriil et al. 2008; Archibald et al. 2016; Gogus et al. 2016) are also accompanied by large amplitude glitches. This strengthens their link to magnetars.

Previously, all the glitches in pulsars and magnetars are spin-up events. In 2012, the magnetar 1E 2259+586 showed a spin-down glitch (i.e. anti-glitch\footnote{In the following, the term ``spin-down glitch'' and ``anti-glitch'' are used interchangeably (Archibald et al. 2017).}), which is also accompanied by burst and outburst (Archibald et al. 2013). This has raised the curiosity what's the physics behind anti-glitches and a variety of models are proposed (Lyutikov 2013; Tong 2014; Katz 2014; Ouyed et al. 2014; Huang \& Geng 2014; Kantor \& Gusakov 2014; Garcia \& Ranea-Sandoval 2015). Normal glitches in pulsars are thought to originate from the interior of the neutron star (Baym et al. 1969; Anderson \& Itoh 1975; Pines \& Alpar 1985). However, since the first anti-glitch in magnetar 1E 2259+586 is accompanied by radiative changes. It is also possible that the anti-glitch is due to processes in the magnetar magnetosphere which can carry away the star's rotational energy (Lyutikov 2013; Tong 2014).

Usually, pulsars and magnetars are assumed to braked down by their magnetic dipole radiation. In this way, the star's characteristic age and magnetic field can be estimated. However, the magnetar showed various kinds of timing activities (Younes et al. 2015;  Camilo et al. 2018; Levin et al. 2019; Tamba et al. 2019; Hu \& Ng 2019; Makishima et al. 2019; Archibald et al. 2020). Many of these timing event are correlated with radiative changes. Variation of particle outflow in the magnetosphere may naturally account for the correlation between radiation and timing in both pulsars (Xu \& Qiao 2001; Kramer et al. 2006; Li et al. 2014; Kou \& Tong 2015) and magnetars (Thompson \& Duncan 1996; Harding et al. 1999; Thompson et al. 2000; Tong et al. 2013a).
This is the wind braking model of pulsars and magnetars. In the case of magnetars, the particle wind luminosity may be order of the X-ray luminosity and much higher than the star's rotational energy loss rate\footnote{When we say rotational energy loss rate, we mean its absolute value $|\dot{E}_{\rm rot}|$, since isolated pulsars and magnetars are slowing down.}: $L_{\rm p} \sim L_{\rm x} \gg \dot{E}_{\rm rot}$. Therefore, the magnetar rotational evolution may be dominated by the particle wind (Harding et al. 1999; Tong et al. 2013a).

During an outburst of a magnetar, the enhanced particle wind can take away some additional rotational of the neutron star and result in net spin-down of the neutron star, i.e. anti-glitch. In this way the first anti-glitch in 1E 2259+586 may be explained (Tong 2014). The wind braking model for anti-glitch has some clear predictions: the anti-glitch should be accompanied by radiative changes, and the duration of the anti-glitch should be relatively long. ``Future observation of one anti-glitch not accompanied by a radiative event or one anti-glitch with a very short time scale can rule out this model'' (Tong 2014).

Recently, another anti-glitch is reported in 1E 2259+586 (Younes et al. 2020). However, this anti-glitch may be radiation-quiet. If it is indeed radiation quiet, then previous magnetospheric models for anti-glitch may be ruled out (Lyutikov 2013; Tong 2014). All previous models may needs fun tuning in order to explain both radiation-loud and radiation-quiet anti-glitches in the same source. However, we will show in below that the 2019 radiation-quiet anti-glitch in 1E 2259+586 is not constraining (section 2). Furthermore, the magnetar 1E 2259+586 is just a typical magnetar. Why it can have so many anti-glitches? This is still unanswered at present. We try to discussion about this in section 3. From the first anti-glitch in 2012 (Archibald et al. 2013), we have more anti-glitch observations from more sources (Archibald et al. 2017; Younes et al. 2020). Considering these observations, we try to discuss a general picture of magnetar anti-glitches in section 4. 

\section{The radiation-quiet anti-glitch in 1E 2259+586 is not constraining}

Younes et al. (2020) reported an anti-glitch in 1E 2259+586 with size $\Delta \nu =-8.3\times 10^{-8} \ \rm Hz$, and $\Delta \nu /\nu =-5.8 \times 10^{-7}$. In their opinion, this anti-glitch has comparable size with the 2012 anti-glitch in 1E 2259+586 $\Delta \nu /\nu \sim -3 \times 10^{-7}$ (Younes et al. 2020). However, this statement is not accurate. The 2019 anti-glitch in 1E 2259+586 has no significant change in $\dot{\nu}$ (Younes et al. 2020). Therefore, the net change in frequency can be viewed as $\Delta \nu/\nu =-5.8\times 10^{-7}$. However, for the 2012 anti-glitch in 1E 2259+586, it has significant changes in both $\nu$ and $\dot{\nu}$ (Archibald et al. 2013). The net change in frequency for the active period is $\Delta \nu =-2.06\times 10^{-7} \ \rm Hz$ and $\Delta \nu /\nu =-1.44 \times 10^{-6}$ (Archibald et al. 2013). Therefore, the 2019 anti-glitch in 1E 2259+586 has an amplitude $2.5$ times smaller than the 2012 anti-glitches. During 2012 anti-glitch, the X-ray flux increased about 100 percent (Archibald et al. 2013). Therefore, the 2019 anti-glitch is expected to have a smaller increase of X-ray flux. Furthermore, during the 2019 anti-glitch only changes in the pulsed flux can be monitored (Younes et al. 2020). This will further weaken the constraining power of the 2019 radiation-quiet anti-glitch.

More quantitative results can be obtained similar to Tong (2014) for the 2012 anti-glitch in 1E 2259+586. The rotational energy loss rate by a strong particle wind is (Harding et al. 1999; Tong et al. 2013a):
\begin{equation}
  \dot{E}_{\rm w} =\dot{E}_{\rm d} \left ( \frac{L_{\rm p}}{\dot{E}_{\rm d}} \right)^{1/2},
\end{equation}
where $\dot{E}_{\rm d}$ is the rotational energy loss rate due to magnetic dipole radiation, $L_{\rm p}$ is the particle wind luminosity. For a typical duration of $\Delta t$ (which can constrained to be about two weeks, i.e. the typical observational interval), an enhanced particle wind will take away some amount of rotational energy of the neutron star and result in net spin-down timing event, i.e. anti-glitch. For 1E 2259+586, in order to result in an anti-glitch with size $|\Delta \nu|$, the total energy and luminosity of the particle outflow are respectively (eq.(4) and (5) in Tong 2014):
\begin{equation}
 \Delta E_{\rm w} =2.3\times 10^{40} I_{45}^2 b_0^{-2} \left( \frac{|\Delta \nu|}{8.3\times 10^{-8} \,\rm Hz} \right)^2
		    \left( \frac{1.2\times 10^6 \,\rm s}{\Delta t} \right) \,\rm erg,
\end{equation}
\begin{equation}
 L_{\rm p} =2\times 10^{34} I_{45}^2 b_0^{-2} \left( \frac{|\Delta \nu|}{8.3\times 10^{-8} \,\rm Hz} \right)^2
		    \left( \frac{1.2\times 10^6 \,\rm s}{\Delta t} \right)^2 \,\rm erg \,s^{-1},
\end{equation}
where $I_{45}$ is the neutron star moment of inertia in units of $10^{45} \ \rm g \ cm^2$, $b_0$ is the star's polar dipole magnetic field in units of the quantum critical field $B_{\rm cr} = 4.4\times 10^{13} \ \rm G$, and the size of the 2019 anti-glitch in 1E 2259+586 is inserted.

For a persistent X-ray flux about $1.7\times 10^{-11} \ \rm erg \ s^{-1} cm^{-2}$ (Younes et al. 2020), and a distance about $3.2 \ \rm kpc$ (Kothes \& Foster 2012), the X-ray luminosity of 1E 2259+586 is about $2.1 \times 10^{34} \ \rm erg \ s^{-1}$. Therefore, the required particle luminosity is comparable with the X-ray luminosity. No significant changes of the X-ray luminosity is required. Furthermore, for the same source 1E 2259+586, its moment of inertial and magnetic field can not change dramatically in 2012 and 2019. Therefore, for a similar duration of the particle wind, the enhanced particle luminosity (and the corresponding X-ray luminosity) is proportional to $|\Delta \nu|^2$. The 2012 anti-glitch is accompanied by an increase of X-ray flux about 100 percent (Archibald et al. 2013). For an anti-glitch with amplitude 2.5 times smaller, the expected increase of X-ray flux should be six ($2.5^2 \approx 6$) times smaller. This will result in an increase of X-ray flux about 20 percent. The pulsed fraction of 1E 2259+586 is about 20 percent (Hu et al. 2019), then the increase of pulsed flux should be about 4 percent. Therefore, the non-detection of significant pulsed flux variation in the 2019 anti-glitch is not constraining.

\section{Why so many anti-glitches in 1E 2259+586?}

Up to now, the magnetar 1E 2259+586 have experienced three glitches (both radiation-loud and radiation-quiet), and three anti-glitches (Dib \& Kaspi 2014; Younes et al. 2020). However, 1E 2259+586 is just a normal magnetar considering its flux and burst etc. Irrespective of the detailed modeling of anti-glitches, we can also ask why 1E 2259+586 can have so many anti-glitches?

The period and period derivative of 1E 2259+586 is $P=7 \ \rm s$ and $\dot{P}=4.8 \times 10^{-13} \ \rm s \ s^{-1}$ (Younes et al. 2020). The period of 1E 2259+586 is normal among the 29 magnetars (Olausen \& Kaspi 2014). However, its period derivative is relatively small compared with normal pulsars, whose period derivative is about $10^{-11} \ \rm s\ s^{-1}$ (Olausen \& Kaspi 2014). This means that 1E 2259+586 has a smaller torque compared with the majority of magnetars. The small period derivative also implies a relative small characteristic magnetic field ($B_{\rm c} =3.2\times 10^{19} \sqrt{P \dot{P}} = 6\times 10^{13} \rm \ G$), a large characteristic age ($\tau_{\rm c} =P/2\dot{P} =2.3\times 10^5 \ \rm yr$), and a low rotational energy loss rate ($\dot{E}_{\rm rot} =4\times 10^{46} \dot{P}/P^3 \ \rm erg \ s^{-1} =6\times 10^{31} \ \rm erg \ s^{-1}$). However, the X-ray luminosity of 1E 2259+586 is relatively high among the magnetar population $L_{\rm x} = 2.1\times 10^{34} \ \rm erg \ s^{-1}$. This means that 1E 2259+586 has a high X-ray luminosity, at the same time its torque is rather small. This may explain why 1E 2259+586 can have so many anti-glitches compared with other magnetars.

The X-ray luminosity of 1E 2259+586 is much higher than its rotational energy loss rate. The X-ray luminosity should come from the magnetic energy release (i.e. magnetar). A very high X-ray luminosity (compared with the rotational energy loss rate) means that its magnetic field is very active. The period-derivative and torque of 1E 2259+586 is relative small. This means that its true magnetic field is relatively small\footnote{Attributing all the spin-down torque to the magnetic dipole spin-down, the characteristic magnetic field is the upper limit of the star's true magnetic field.}. The particle wind luminosity in the magnetosphere may be comparable with the X-ray luminosity $L_{\rm p} \sim L_{\rm x}$. Therefore, the rotational evolution of 1E 2259+586 may be dominated by the particle wind. A small glitch may trigger some kind of X-ray flux enhancement. The particle wind may also be enhanced at the same time. An enhanced particle wind may take away additional rotational energy of the neutron star. The final output may be an anti-glitch.

\begin{figure}
\centering
\includegraphics[width=0.5\textwidth]{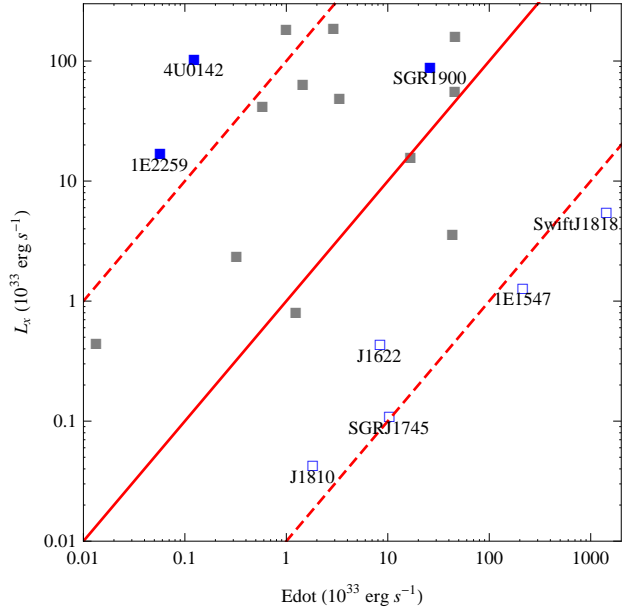}
\caption{X-ray luminosity vs. rotational energy loss rate for 24 magnetars. The blue squares are the three magnetars with anti-glitches. The empty blue squares are the radio emitting magnetars. The rest magnetars are marked in grey squares. The solid red line means the the X-ray luminosity and rotational energy are equal. The upper and lower red dashed line is for $L_{\rm x}=10^{2} \ \dot{E}_{\rm rot}$ and $L_{\rm x}=10^{-2} \ \dot{E}_{\rm rot}$, respectively.}
\label{gLxEdot}
\end{figure}

To characterize the above picture more quantitatively, the X-ray luminosity and rotational energy loss rate of 24 magnetar is shown in \ref{gLxEdot}. From the magnetar cataloge (Olausen \& Kaspi 2014; www.physics.mcgill.ca/$\sim$pulsar/magnetar/main.html
), 23 magnetar have both period and period derivative. Among them, one magnetar SGR 1935+2154 has no persistent X-ray luminosity reported.
It persistent X-ray flux is added from Hu et al. (2019). The fifth radio-emitting magnetar Swift J1818.0-1607 is added from Esposito et al. (2020). Therefore, this left us a total of 24 magnetars with both X-ray luminosity\footnote{In several cases, only the upper limit on the quiescent X-ray luminosity is reported (e.g., for Swift J1818.0-1607, Espositio et al. 2020). We use the reported upper limit as an approximation for the quiescent X-ray luminosity.} and rotational energy loss rate.  It has been noted previously that magnetars with low X-ray luminosity are more like to have radio emission (Rea et al. 2012; Tong et al. 2013b). Up to know, three magnetars are observed to have anti-glitches (Archibald et al. 2017; Younes et al. 2020): 1E 2259+586 (three anti-glitches), 4U 0142+61 (two anti-glitches), SGR 1900+14 (one anti-glitch). We note that all these three anti-glitchers have an X-ray luminosity much higher than their rotational energy loss rate $L_{\rm x} \gg \dot{E}_{\rm rot}$. Furthermore, the two most prolific anti-glitchers (1E 2259+586 and 4U 0142+61) have an X-ray luminosity 100 times higher than their rotational energy loss rate $L_{\rm x} \ge 100 \dot{E}_{\rm rot}$. The X-ray luminosity and rotational energy loss rate represent the two driving forces for the putative neutron star: magnetic energy vs. rotational energy. Therefore, the anti-glitchers and radio-emitter of magnetars occupy opposite corner on the $L_{\rm x}$-$\dot{E}_{\rm rot}$ diagram.

\begin{figure}
\centering
\includegraphics[width=0.5\textwidth]{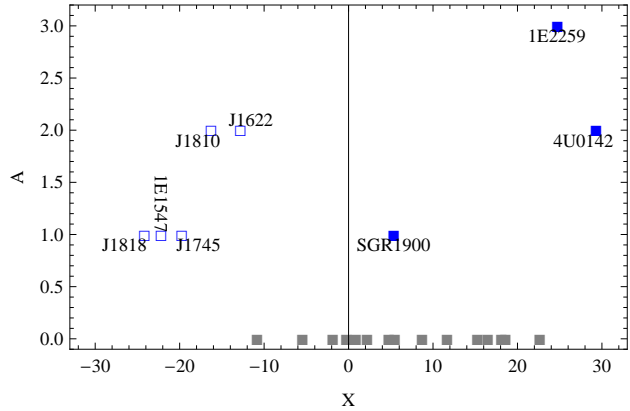}
\caption{Activity parameter vs. the X-ray luminosity parameter. For positive X-ray luminosity parameter (X-ray luminosity higher than the rotational energy loss rate), the magnetars are more likely to have anti-glitches, blue squares. For negative X-ray luminosity parameter (X-ray luminosity smaller than the rotational energy loss rate), the magnetars are more likely to have radio emissions, empty blue squares. The rest of the magnetars are marked in gray squares, with activity parameter $A=0$, i.e. no anti-glitch or radio emission detected.}
\label{gAX}
\end{figure}

To characterize the above argument more quantitatively, we define the following X-ray luminosity parameter
\begin{equation}
  X= 10 \log_{10} \left( \frac{L_{\rm x}}{\dot{E}_{\rm rot}} \right).
\end{equation}
For an $X=20$ means the X-ray luminosity is $100$ times the rotational energy loss rate, an $X=-20$ means the X-ray luminosity is $10^{-2}$ times the rotational energy loss rate etc. To characterize the anti-glitch or radio emitting activity, we introduce the activity parameter $A$: which is the number of detected anti-glitches, or the number of active radio emitting epoch. For example, for 1E 2259+586 which has three anti-glitches, its activity parameter $A=3$. For the radio emitting magnetar XTE J1810-197 and PSR J1622-4950 which showed two active radio emitting epoch, their activity parameters are $A=2$ (Dai et al. 2019; Levin et al. 2019; Camilo et al. 2018). SGR 1935+2154 has an X-ray luminosity comparable with its rotational energy loss rate (Hu et al. 2019). It is not considered as a radio emitter. Its radio burst is different from the previous pulsed radio emission of the five magnetars (Andersen et al. 2020; Bochenek et al. 2020). The activity parameter vs. the X-ray luminosity parameter is shown in figure \ref{gAX}. The correlation between anti-glitch/radio emission and X-ray luminosity parameter is clearly seen. And the anti-glitch and radio emitting magnetars are on the opposite end of the X-ray luminosity parameter. From figure \ref{gAX}, another feature can be seen. In both sub-domain ($X>0$ and $X<0$), the activity parameter seems to be positively correlated with the X-ray luminosity parameter.
Since the X-ray luminosity are due to magnetic energy release, this may indicate that both the radio-emission and anti-glitch are driven by the magnetic energy release. Considering that the current number of radio-emitter and anti-glitchers are rather small, the above interpretation should be taken with care and need confirmation in the future.


\section{The general picture of magnetar anti-glitches}

With more anti-glitches discovered (Archibald et al. 2017; Younes et al. 2020), a general picture of magnetar anti-glitch may be obtained.
\begin{itemize}
  \item For a glitch in a magnetar, if it can trigger large scale magnetic field reconfiguration (Beloborodov 2009; Zhou et al. 2014; Tong 2019), a period of significant magnetic release may be initiated. Otherwise, the glitch will be radiation-quiet, like that in normal pulsars.
      Therefore, there can be both radiation-quiet and radiation-loud glitches in magnetars.
  \item The magnetic energy release rate should be larger than or comparable with the rotational energy loss rate in order for significant magnetar-like activity to be seen from it (Tong \& Huang 2020). Therefore, only magnetars with relative high magnetic field can we see its magnetic activities.
      These include high magnetic field pulsars and normal magnetars.
  \item A large amount of rotational energy should be taken away in order to make the glitch (which trigger the outburst) to become an anti-glitch (Tong 2014; Tong \& Huang 2020). Therefore, an outburst of magnetars much be accompanied by either glitch or anti-glitches. Especially, in this magnetospheric scenario, every anti-glitch should be radiation-loud. The 2019 radiation-quiet anti-glitch in 1E 2259+586 is not constraining as shown above. Future observations of a radiation-quiet anti-glitch with an amplitude larger than or comparable to the 2012 one can constrain the magnetospheric models for anti-glitches.
\end{itemize}


The high magnetic field pulsar PSR J1846-0258 have nonthermal X-ray spectrum and a pulsar wind nebula (Gavriil et al. 2008; Kuiper \& Hermsen 2009). Therefore, it is similar to PSR J1119-6127, only that the radio beam of PSR J1846-0258 are not pointing toward us. Both of these sources have shown magnetar-like activities, triggered by a glitch which finally become a spin-down glitch (Livingstone et al. 2011; Dai et al. 2018; Archibald et al. 2018).
The magnetar outburst may accompanied by a particle outflow. This outflow may also be responsible for the spin-glitch in both of these sources. Similar to that in magnetar 1E 2259+586. The persistent state X-ray and/or radio emission of PSR J1846-0258/PSR J1119-6127 are though to be rotation-powered. It is not an indicator of magnetic activity. Therefore, it is not shown on figure \ref{gLxEdot} and \ref{gAX}.

The X-ray luminosity of magnetars should come from the magnetic energy release. However, the radio emission of magnetars can be powered by either the rotational energy or the magnetic energy. Its origin is still unknown (Rea et al. 2012; Tong et al. 2013b). From figure \ref{gAX}, the positive correlation between radio activity parameter and X-ray luminosity parameter may give us some hint that the radio emission is powered by the magnetic energy. The coherent radio emission of pulsars and magnetar may be similar. However, the radio emission mechanism of normal pulsars is undetermined. This is also the bottle-neck of current magnetar radio emission researches.

In conclusion, the 2019 radiation-quiet anti-glitch in 1E 2259+586 is not constraining, since its it has a smaller amplitude. Magnetars with high X-ray luminosity and low rotational energy loss rate are more likely to have anti-glitches. Anti-glitchers and radio-emitters lie at opposite end about the X-ray luminosity distribution. More physical modeling are needed in the future.

\section*{Acknowledgments}
H.Tong is supported by NSFC (11773008).





\begin{thebibliography}{99}

\bibitem{Andersen2020}
Andersen, B. C., Bandura, K. M., Bhardwaj, M., et al., 2020, arXiv:2005.10324

\bibitem{Anderson1975}
Anderson, P. W., \& Itoh, N., 1975, Nature, 256, 25

\bibitem{Archibald2013}
Archibald, R. F., Kaspi, V. M., Ng, C. Y., et al., 2013, Nature, 497, 591

\bibitem{Archibald2016}
Archibald, R. F., Kaspi, V. M., Tendulkar, S. P., et al., 2016, ApJ, 829, L21

\bibitem{Archibald2017}
Archibald, R. F., Kaspi, V. M., Scholz, P., et al., 2017, ApJ, 834, 163

\bibitem{Archibald2018}
Archibald, R. F., Kaspi, V. M., Tendulkar, S. P., \& Scholz, P., 2018, ApJ, 869, 180	

\bibitem{Archibald2020}
Archibald, R. F., Scholz, P., Kaspi, V. M., et al., 2020, arXiv:2001.06450

\bibitem{Beloborodov2009}
Beloborodov, A. M., 2009, ApJ, 703, 1044

\bibitem{Baym1969}
Baym, G., Pethick, C., Pines, D., \& Ruderman, M., 1969, Nature, 224, 872

\bibitem{Bochenek2020}
Bochenek, C. D., Ravi, V., Belov, K. V., et al., 2020, arXiv:2005.10828

\bibitem{Camilo2018}
Camilo, F., Scholz, P., Serylak, M., et al., 2018, ApJ, 856, 180

\bibitem{Dai2018}
Dai, S., Johnston, S., Weltevrede, P., et al., 2018, MNRAS, 480, 3584

\bibitem{Dai2019}
Dai, S., Lower, M. E., Bailes, M., et al., 2019, ApJL, 874, L14

\bibitem{Dib2014}
Dib, R., \& Kaspi, V. M., 2014, ApJ, 784, 37

\bibitem{Esposito2020}
Esposito, P., Rea, N., Borghese, A., et al., 2020, arXiv:2004.04083

\bibitem{Gavriil2008}
Gavriil, F. P., Gonzalez, M. E., Gotthelf, E. V., et al., 2008, Science, 319, 1802

\bibitem{Gogus2016}
Gogus, E., Lin, L., Kaenko, Y., et al., 2016, ApJ, 829, L25

\bibitem{Gracia2015}
Gracia, F., \& Ranea-Sandoval, I. F., 2015, MNRAS, 449, L73

\bibitem{Harding1999}
Harding, A. K., Contopoulos, I., \& Kazanas, D. 1999, ApJ, 525, L125

\bibitem{Hu2019}
Hu, C. P., \& Ng, C. Y., 2019, AN, 340, 340

\bibitem{Hu2019b}
Hu, C. P., Ng, C. Y., \& Ho, W. C. G., 2019, MNRAS, 485, 4274

\bibitem{Huang2014}
Huang, Y. F., \& Geng, J. J., 2014, ApJL, 782, L20

\bibitem{Kantor2014}
Kantor, E. M., \& Gusakov, M. E., 2014, ApJL, 797, L4

\bibitem{Katz2014}
Katz, J. I., 2014, ApSS, 349, 611

\bibitem{Kothes2012}
Kothes, R., \& Foster, T. 2012, ApJL, 746, L4,

\bibitem{Kou2015}
Kou, F. F., \& Tong, H., 2015, MNRAS, 450, 1990

\bibitem{Kramer2006}
Kramer, M., Lyne, A. G., O'Brien, J. T., et al., 2006, Science, 312, 549	

\bibitem{Kuiper2009}
Kuiper, L., \& Hermsen, W., 2009, A\&A, 501, 1031

\bibitem{Levin2019}
Levin, L., Lyne, A. G., Desvignes, G., et al., 2019, MNRAS, 488, 5251

\bibitem{Li2014}
Li, L., Tong, H., Yan, W. M., et al., 2014, ApJ, 788, 16

\bibitem{Livingstone2011}
Livingston M. A., Ng C. Y., Kaspi V. M., et al., 2011, ApJ, 730, 66

\bibitem{Lyne2012}
Lyne, A. G., \& Graham-Smith, F., 2012, Pulsar astronomy, Cambridge University, Cambridge

\bibitem{Lyutikov2013}
Lyutikov, M., 2013, arXiv:1306.2264

\bibitem{Makishima2019}
Makishima, K., Murakami, H., Enoto, T., \& Nakazawa, K., 2019, PASJ, 71, 15

\bibitem{Olausen2014}
Olausen, S. A., \& Kaspi, V. M., 2014, ApJS, 212, 6

\bibitem{Ouyed2014}
Oyued, R., Leahy, D., \& Koning, N., 2014, ApSS, 352, 715

\bibitem{Pines1985}
Pines, D., \& Alpar, M. A., 1985, Nature, 316, 27

\bibitem{Rea2012}
Rea, N., Pons, J. A., Torres, D. F., \& Turolla, R. 2012, ApJ, 748, L12

\bibitem{Tamba2019}
Tamba, T., Bamba, A., Odaka, H., \& Enoto, T., 2019, PASJ, 71, 90

\bibitem{Thompson1996}
Thompson, C., \& Duncan, R. C., 1996, ApJ, 473, 322

\bibitem{Thompson2000}
Thompson, C., Duncan, R. C., Woods, P. M., et al. 2000, ApJ, 543, 340

\bibitem{Tong2013a}
Tong, H., Xu, R. X., Song, L. M., \& Qiao, G. J., 2013a, ApJ, 768,144

\bibitem{Tong2013b}
Tong, H., Yuan, J. P., \& Liu, Z. Y., 2013b, RAA, 13, 835

\bibitem{Tong2014}
Tong, H., 2014, ApJ, 784, 86

\bibitem{Tong2019}
Tong, H., 2019, MNRAS, 489, 3769

\bibitem{Tong2020}
Tong, H., \& Huang, L., 2020, arXiv:2005.11281

\bibitem{Xu2001}
Xu, R. X., \& Qiao, G. J., 2001, ApJL, 561, L85

\bibitem{Younes2015}
Younes, G., Kouveliotou, C., \& Kaspi, V. M., 2015, ApJ, 809, 165

\bibitem{Younes2020}
Younes, G., Ray, P. S., Baring, M. G., et al., 2020, arXiv:2006.04854

\bibitem{Zhou2014}
Zhou, E. P., Lu, J. G., Tong, H., \& Xu, R. X., 2014, MNRAS, 443, 2705

\end{thebibliography}


%


\end{document}